\documentclass[
reprint,
superscriptaddress,
 amsmath,amssymb,
 aps,
prb,
floatfix,
]{revtex4-2}

\usepackage{graphicx}
\usepackage{dcolumn}
\usepackage{bm}
\usepackage{booktabs} 
\usepackage{xcolor}
\usepackage[normalem]{ulem}
\usepackage{url}

\counterwithout{figure}{section}          

\begin{document}

\preprint{APS/123-QED}

\title{Thermal Hall conductivity in the strongest cuprate superconductor : \\ Estimate of the mean free path in the trilayer cuprate \(\text{HgBa}_2\text{Ca}_2\text{Cu}_3\text{O}_{8 + \delta} \)}

\author{Munkhtuguldur Altangerel}\email{munkhtuguldur.altangerel@usherbrooke.ca}
 \affiliation{Institut Quantique, Département de physique \& RQMP, Université de Sherbrooke, Sherbrooke, Québec, Canada J1K 2R1}
 \affiliation{LNCMI-EMFL, CNRS UPR3228, Univ. Grenoble Alpes, Univ. Toulouse, INSA-T, Grenoble and Toulouse, France}
 \affiliation{Université de Sherbrooke – CNRS / IRL Frontières Quantiques, Sherbrooke, Canada}
\author{Quentin Barthélemy}
 \affiliation{Institut Quantique, Département de physique \& RQMP, Université de Sherbrooke, Sherbrooke, Québec, Canada J1K 2R1}
\author{Étienne Lefrançois}
 \affiliation{Institut Quantique, Département de physique \& RQMP, Université de Sherbrooke, Sherbrooke, Québec, Canada J1K 2R1}
\author{Jordan Baglo}
 \affiliation{Institut Quantique, Département de physique \& RQMP, Université de Sherbrooke, Sherbrooke, Québec, Canada J1K 2R1}
\author{Manel Mezidi}
 \affiliation{Institut Quantique, Département de physique \& RQMP, Université de Sherbrooke, Sherbrooke, Québec, Canada J1K 2R1}
 \affiliation{Université de Sherbrooke – CNRS / IRL Frontières Quantiques, Sherbrooke, Canada}
 \affiliation{Université Paris-cité, Laboratoire Matériaux et Phénomènes Quantiques, CNRS (UMR 7162), 75013 Paris, France}
\author{Gaël Grissonnanche}
 \affiliation{Laboratoire des Solides Irradiés, CEA/DRF/lRAMIS, CNRS, École Polytechnique, Institut Polytechnique de Paris, F-91128 Palaiseau, France}
\author{Ashvini Vallipuram}
 \affiliation{Institut Quantique, Département de physique \& RQMP, Université de Sherbrooke, Sherbrooke, Québec, Canada J1K 2R1}
\author{Emma Campillo}
 \affiliation{Institut Quantique, Département de physique \& RQMP, Université de Sherbrooke, Sherbrooke, Québec, Canada J1K 2R1}
\author{Anne Forget}
 \affiliation{Université Paris-Saclay, CEA, CNRS, SPEC, 91191, Gif-sur-Yvette, France}
\author{Dorothée Colson}
 \affiliation{Université Paris-Saclay, CEA, CNRS, SPEC, 91191, Gif-sur-Yvette, France}
\author{Ruixing Liang}
 \affiliation{Department of Physics and Astronomy, University of British Columbia, Vancouver, British Columbia, Canada}
 \affiliation{Canadian Institute for Advanced Research, Toronto, Ontario, Canada M5G 1M1}
\author{D. A. Bonn}
 \affiliation{Department of Physics and Astronomy, University of British Columbia, Vancouver, British Columbia, Canada}
 \affiliation{Canadian Institute for Advanced Research, Toronto, Ontario, Canada M5G 1M1}
\author{W. N. Hardy}
 \affiliation{Department of Physics and Astronomy, University of British Columbia, Vancouver, British Columbia, Canada}
 \affiliation{Canadian Institute for Advanced Research, Toronto, Ontario, Canada M5G 1M1}
\author{Cyril Proust}
 \affiliation{LNCMI-EMFL, CNRS UPR3228, Univ. Grenoble Alpes, Univ. Toulouse, INSA-T, Grenoble and Toulouse, France}
 \affiliation{Université de Sherbrooke – CNRS / IRL Frontières Quantiques, Sherbrooke, Canada}
\author{Louis Taillefer}\email{louis.taillefer@usherbrooke.ca}
 \affiliation{Institut Quantique, Département de physique \& RQMP, Université de Sherbrooke, Sherbrooke, Québec, Canada J1K 2R1}
  \affiliation{Université de Sherbrooke – CNRS / IRL Frontières Quantiques, Sherbrooke, Canada}
 \affiliation{Canadian Institute for Advanced Research, Toronto, Ontario, Canada M5G 1M1}

\date{\today}
\

\begin{abstract}

The thermal Hall conductivity of the trilayer cuprate HgBa\textsubscript{2}Ca\textsubscript{2}Cu\textsubscript{3}O\textsubscript{8+$\delta$} (Hg1223) – the superconductor with the highest critical temperature $T_c$ at ambient pressure – was measured at temperatures down to 2 K for three dopings in the underdoped regime (\(p = 0.09, 0.10, 0.11\)). By combining a previously introduced simple model and prior theoretical results, we derive a formula for the inverse mean free path, \(1 / \ell\), which allows us to estimate the mean free path of $d$-wave quasiparticles in Hg1223 below $T_c$. We find that \(1 / \ell\) grows as $T^3$, in agreement with the theoretical expectation for a clean \textit{d}-wave superconductor. Measurements were also conducted on the single layer mercury-based cuprate HgBa\textsubscript{2}CuO\textsubscript{6+$\delta$} (Hg1201), revealing that the mean free path in this compound is roughly half that of its three-layered counterpart at the same doping \((p = 0.10)\). This observation is attributed to the protective role of the outer planes in Hg1223, which results in a more pristine inner plane. We also report data in an ultraclean crystal of YBa\textsubscript{2}Cu\textsubscript{3}O\textsubscript{\textit{y}} (YBCO) with full oxygen content \((p = 0.18)\), believed to be the cleanest of any cuprate, and find that \(\ell\) is not longer than in Hg1223.

\end{abstract}

\maketitle

\section{\label{sec:level1}INTRODUCTION}

The mean free path is a key property of electrons in metals. The most useful and simplest way to estimate the electronic mean free path $\ell_{\mathrm{n}}$ is via the electrical resistivity $\rho$ measured in the normal state, without superconductivity. In particular, the residual value at $T \to 0$, $\rho_0$, yields the elastic mean free path $\ell_{\mathrm{n0}}$, via the Drude formula
\[
\rho_0 = \left(\frac{m^*}{n e^2}\right)\left(\frac{1}{\tau_{\mathrm{n0}}}\right),
\]
with $\ell_{\mathrm{n0}} = v_{\mathrm{F}} \tau_{\mathrm{n0}}$, where $v_{\mathrm{F}}$ is the Fermi velocity, $\tau_{\mathrm{n0}}$ is the scattering time at $T \to 0$, $m^*$ is the effective mass, $n$ is the density of charge carriers and $e$ is the electron charge. The mean free path $\ell_{\mathrm{n0}}$ and the scattering rate $1/\tau_{\mathrm{n0}}$ are quantitative measures of the scattering of electrons on defects in a sample in the normal state.

In $d$-wave superconductors, impurity scattering has a strong impact on superconducting properties because it easily breaks pairs. The strength of pair breaking is quantified by the ratio \(\frac{\hbar/\tau_{\mathrm{n0}}}{k_{\mathrm{B}} T_{\mathrm{c}}}\) (with $\hbar$ the reduced Planck constant $h/2\pi$, $k_{\mathrm{B}}$ the Boltzmann constant and $T_{\mathrm{c}}$ the superconducting critical temperature). A ratio approaching 1 is expected to produce visible effects \cite{Hirschfeld1993,Park1997}, such as a lower $T_{\mathrm{c}}$, a lower superfluid density and a deviation from the universal thermal conductivity at $T \to 0$ \cite{maki1995}.

In cuprate superconductors, it has not been possible to estimate the electron mean free path in most samples because their robust superconductivity makes it difficult to measure $\rho$ down to low temperatures. Measurements in high magnetic fields have provided some $\rho_0$ values in samples that have a relatively low upper critical field $H_\text{c2}$, typically at dopings near $p = 0.12$ or above $p \simeq 0.2$ \cite{Grissonnanche2014}. For example, in overdoped \(\text{Tl}_2\text{Ba}_2\text{CuO}_{6+\delta}\)  (Tl2201) at $p = 0.29$ ($T_{\mathrm{c}} = 15$~K, \(H_\text{c2}(0) = 15\) T), $\rho_0 \simeq 6$~$\mu \Omega\cdot\text{cm}$ \cite{proust202}, and in overdoped \(\text{Bi}_2\text{Sr}_2\text{CaCu}_2\text{O}_{8+\delta}\) (Bi2212) at $p = 0.23$ ($T_{\mathrm{c}} = 50$~K, \(H_\text{c2}(0) = 50\) T), $\rho_0 \simeq 20$~$\mu \Omega\cdot\text{cm}$ \cite{Legros2019}. But in neither of these materials do we have any idea of $\rho_0$ in samples near optimal doping, since $H_\text{c2} \simeq 150$~T at such dopings \cite{Grissonnanche2014}.

In cuprates with a lower $H_\text{c2}$, resistivity measurements in pulsed fields up to 60--90 T can yield values for $\rho$ in the $T \to 0$ limit across the full doping range. In overdoped \(\text{La}_{2-x}\text{Sr}_x\text{CuO}_4\) (LSCO) with $p = 0.23 - 0.24$ \cite{Cooper2009,Ataei2022} and \(\text{La}_{1.6-x}\text{Nd}_{0.4}\text{Sr}_x\text{CuO}_4\) (Nd-LSCO) with $p = 0.24$ \cite{Daou2009,Collignon2017}, this typically gives $\rho_0 \simeq 20$~$\mu \Omega\cdot\text{cm}$. However, in the underdoped regime, it is difficult to deduce a mean free path from the much larger values of $\rho$ in the $T \to 0$ limit, e.g. $\rho_0 \simeq 400$~$\mu \Omega\cdot\text{cm}$ in LSCO at $p \simeq 0.13$ \cite{Laliberte-arXiv2016}, because we have little knowledge of the Fermi surface in that doping range, where pseudogap, spin glass phase and charge order prevail and transform it profoundly \cite{Proust-Taillefer2019,frachet2020a,campbell2024}.

There is clearly a need for another way to measure the electronic scattering rate and the mean free path in cuprates. Two decades ago, Zhang \textit{et al.} \cite{Zhang2001} proposed an approach that has not been exploited much since. It is based on a measurement of the thermal Hall conductivity $\kappa_{{xy}}$, which can be performed below $T_{\mathrm{c}}$ in modest magnetic fields. The benefit of employing $\kappa_{{xy}}$ as opposed to thermal conductivity $\kappa_{{xx}}$ is that in $\kappa_{{xx}}$, phonons contribute significantly alongside electrons, namely $\kappa_{{xx}} = \kappa_{\mathrm{qp}} + \kappa_\mathrm{ph}$. By contrast, in $\kappa_{{xy}}$, the contribution of phonons is negligible at low fields, allowing one to extract the electronic mean free path. A simple model then yields an estimate of $\ell_{\mathrm{s}}$, the mean free path of quasiparticles in the superconducting state. In ultra-clean YBCO at $p = 0.18$, Zhang \textit{et al.} found that $\ell_{\mathrm{s}} \simeq 1$~$\mu$m at $T \to 0$ and $H \to 0$ \cite{Zhang2001}. 

However, this model requires a knowledge of the specific heat of nodal quasiparticles. Concurrently, Vekhter \textit{et al.} developed a theoretical formula for the specific heat of nodal quasiparticles \cite{vekhter201}. By substituting their formula into the model proposed by Zhang \textit{et al.} \cite{Zhang2001}, one can derive an expression for the electronic mean free path that requires merely the thermal Hall conductivity \(\kappa_{{xy}}\), the average distance between CuO\(_2\) layers \(d\), the Fermi wavevector \(k_{\mathrm{F}}\) and the gap velocity \(v_{\mathrm{\Delta}}\). $v_{\mathrm{\Delta}}$ is the slope of the $d$-wave gap at the node, which can either be measured by ARPES \cite{Vishik2010,Vishik2012} or accessed via the thermal conductivity in the $T \to 0$ limit \cite{Taillefer1997,Durst2000,sutherland2003}.

In this Article, we report measurements of $\kappa_{{xy}}$ in three different cuprates: trilayer Hg1223 at $p = 0.09, 0.10$ and 0.11; single-layer Hg1201 at $p = 0.10$; bilayer YBCO at $p=0.18$. 
In the superconducting state, we find that \( 1 / \ell_{\mathrm{s}} \) displays a \( T^3 \) dependence, as expected theoretically for a clean \( d \)-wave superconductor \cite{walker2000, dahm2005, duffy2001, Quinlan1994}.
We obtain the residual value of the mean free path, \( \ell_{\mathrm{s0}} \) at \( T \to 0 \), for each sample, allowing us to compare amongst cuprates. We find that our Hg1223 samples are as clean as the cleanest YBCO samples.

\section{\label{sec:level2}METHODS}
\subsection{Samples}

High-quality single crystals of Hg1223 with dopings \(p = 0.09\), \(p = 0.10\), and \(p = 0.11\) were grown using the self-flux technique \cite{loret2017}. 
Doping levels were determined using the empirical relationship \cite{PRESLAND199195}:
\[
1 - T_{\mathrm{c}} / T_{\mathrm{c},\text{max}} = 82.6(p - 0.16)^2,
\]  

where \(T_{\mathrm{c}}\) denotes the onset of the superconducting transition and \(T_{\mathrm{c},\text{max}}\), the critical temperature of optimally doped Hg1223, is 135 K. The \(p = 0.09\) sample had \(T_{\mathrm{c}} = 78 \ \text{K}\), the \(p = 0.10\) sample had \(T_{\mathrm{c}} = 95 \ \text{K}\), and the \(p = 0.11\) sample had \(T_{\mathrm{c}} = 112 \ \text{K}\).

A single high-purity crystal of underdoped Hg1201 (\( p = 0.10 \), \( T_{\mathrm{c}} = 76 \) K) was measured, prepared as described in \cite{PhysRevB.78.054518,zhao2006}. The doping level was determined based on the \( T_{\mathrm{c}}(p) \) relationship for Hg1201 established in \cite{PhysRevB.63.024504}.

Single crystals of YBCO with \(p = 0.18\) were grown by flux growth, as described in \cite{liang2006}. Our sample was a single detwinned crystal with an oxygen content \(y = 6.998\), corresponding to \(T_{\mathrm{c}} = 90.5 \ \text{K}\).

All samples were prepared as rectangular platelets, with gold sputtered contacts and subsequent silver paint applied for electrical measurements. 
The typical dimensions of the samples are \(300-1000 \times 300 \, \mu\text{m}\), with a thickness of $\sim 100~\mu$m.

The inter-plane distance \(d\) represents the average separation between CuO\(_2\) planes. For Hg1201, \(d = 9.5 \, \text{\AA}\), corresponding to the distance between CuO\(_2\) planes. For YBCO, \(d = 5.8 \, \text{\AA}\), the average separation between CuO\(_2\) planes in the unit cell. In the case of Hg1223, there are two possible ways to define \(d\): \(d = \frac{c}{3}\) if we treat all three planes as equivalent, where \(c = 15.86 \, \text{\AA}\), or \(d = c\) if we consider only the distance between the inner planes (see discussion below). The values of \( T_{\mathrm{c}}\) and \(d\) are given in Table \ref{tab:sample_properties}.

\begin{table}
\centering
\setlength{\tabcolsep}{6pt} 
\renewcommand{\arraystretch}{1.2} 
\begin{tabular}{cccccc}
\toprule
\toprule

Compound & \(T_{\mathrm{c}}\) (K) & \(p\) & \(d\) (\AA) & \(\ell_{\mathrm{s0}}\) (\(\text{\AA}\))  \\
\midrule

Hg1223 & 78   & 0.09  & 5.3 / 16 & 1180 $\pm$ 170  \\
Hg1223 & 95   & 0.10  & 5.3 / 16 & 1590 $\pm$ 230\\
Hg1223 & 112  & 0.11  & 5.3 / 16 & 2760 $\pm$ 390 \\
Hg1201 & 76   & 0.10  & 9.5        & 1040 $\pm$ 150 \\
YBCO   & 90.5 & 0.18  & 5.8        & 3530 $\pm$ 500 \\

\bottomrule
\bottomrule
\end{tabular}
\caption{\textbf{Properties of the measured samples}, including the superconducting transition temperature \(T_{\mathrm{c}}\), hole doping level \(p\), inter-plane distance \(d\), and mean free path \(\ell_{\mathrm{s0}}\) at \(T \to 0\) and \(B = 0.5\) T. The error bars on \(\ell_{\mathrm{s0}}\) include uncertainty from approximation, fitting and geometrical factors. Here, \(d\) represents the inter-plane distance, with \(d = 5.3 \, \text{\AA}\) or \(d = 16 \, \text{\AA}\) for Hg1223, depending on whether it is calculated as \(c/3\) or \(c\), respectively. The \(\ell_{\mathrm{s0}}\) values for Hg1223 correspond to the case \(d = c/3\) (with all planes considered equivalent).}

\label{tab:sample_properties}
\end{table}

\subsection{Thermal transport measurement}
\begin{figure*}
\centering
\includegraphics[width=1\textwidth]{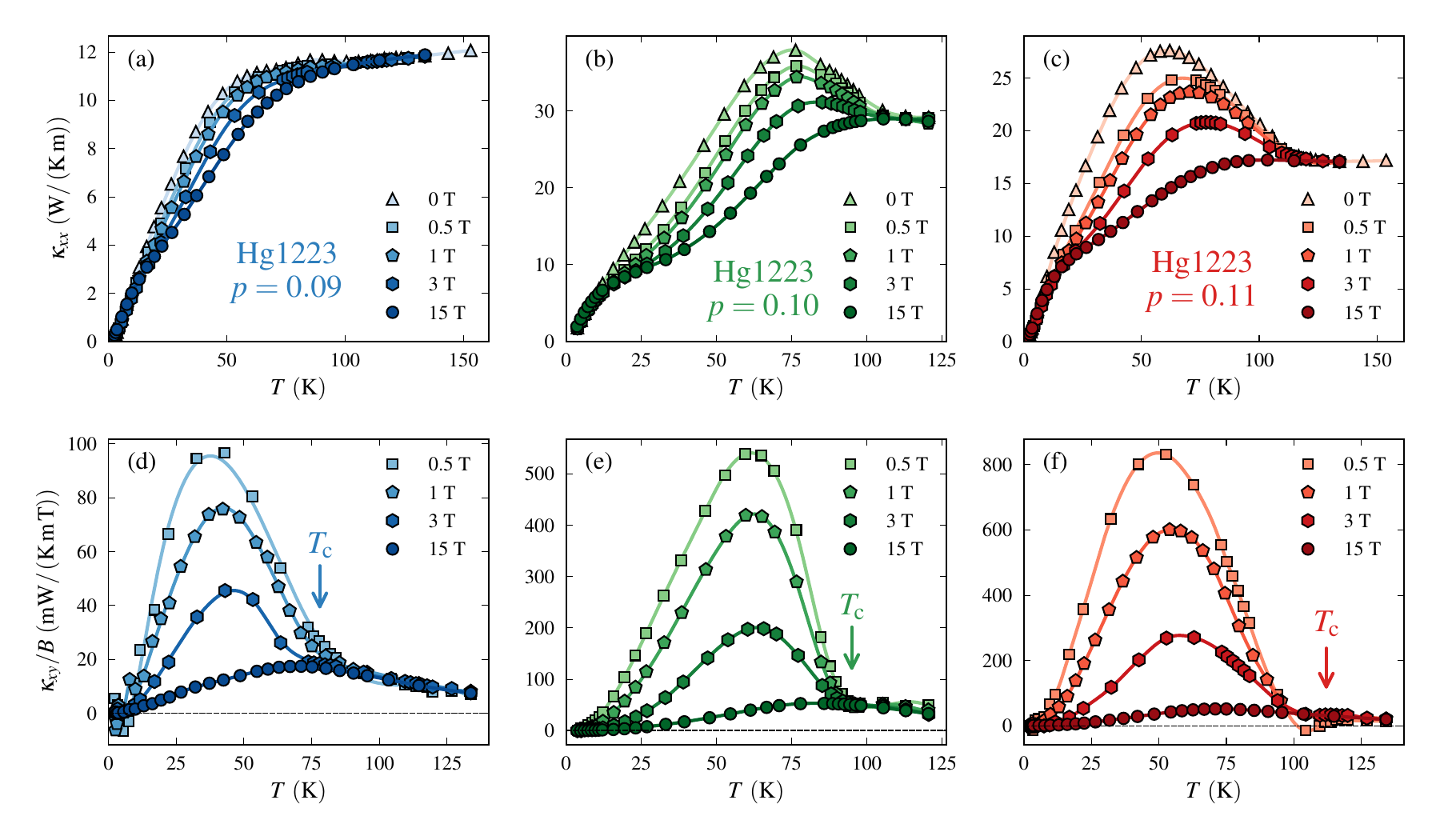}
\caption{\textbf{Thermal conductivity of Hg1223.} 
Longitudinal thermal conductivity \(\kappa_{{xx}}\)  and thermal Hall conductivity normalized by field \(\kappa_{{xy}}/B\) for Hg1223 (\(p = 0.09, 0.10, 0.11\)) as a function of temperature \(T\) in magnetic fields of 0 T (triangles), 0.5 T (squares), 1 T (pentagons), 3 T (hexagons), and 15 T (circles), with darker shades indicating higher fields. Lines serve as guides to the eye, while markers represent the data points. Panels (a)–(c) show \(\kappa_{{xx}}\): (a) Hg1223 \(p = 0.09\) (blue); (b) Hg1223 \(p = 0.10\) (green); (c) Hg1223 \(p = 0.11\) (red). Panels (d)–(f) show \(\kappa_{{xy}}/B\). The arrows indicate the location of the superconducting transition.}
\label{Fig_Hg1223_kxx_kxy}
\end{figure*}
The measurement of thermal conductivity \( \kappa_{{xx}} \) involves applying a heat power \( \dot{Q}_x \) along the \( x \) axis of the sample, causing a longitudinal temperature difference \( \Delta T_{{x}} = T^{+} - T^{-} \). Eq.~(\ref{eq:kxx}) provides the expression for \( \kappa_{{xx}} \):

\begin{equation}
    \kappa_{{xx}} = \frac{\dot{Q}_x }{\Delta T_{{x}}} \left( \frac{L}{wt} \right)
    \label{eq:kxx}
\end{equation} 

where \( w \) is the width of the sample, \( t \) is its thickness, and \( L \) is the distance between \( T^{+} \) and \( T^{-} \). When a magnetic field is applied along the \( z \) axis, a transverse temperature difference \( \Delta T_{{y}} \) can develop along the \( y \) axis. Eq.~(\ref{eq:kxy}) defines the thermal Hall conductivity \( \kappa_{{xy}} \):

\begin{equation}
    \kappa_{{xy}} = -\kappa_{{yy}} \frac{\Delta T_{{y}}}{\Delta T_{{x}}} \frac{L}{w}
    \label{eq:kxy}
\end{equation}

with \( \kappa_{{yy}} \) representing the thermal conductivity in the \( {y} \) direction. This expression is an approximation that holds when \( \kappa_{{xy}} / \kappa_{{xx}} \ll 1 \), a condition that is satisfied for all our measurements. In a tetragonal system such as Hg1223 and Hg1201, \( \kappa_{{yy}}\) is equal to \(\kappa_{{xx}} \). The temperature differences \( \Delta T_{{y}} \) and \( \Delta T_{{x}} \) were measured using type E thermocouples (chromel constantan) in a steady-state method in a fixed magnetic field \(B\). This choice was based on the weak-field dependence of type-E thermocouples within the explored temperature and magnetic-field range, offering a better sensitivity than resistive Cernox sensors at high temperatures.

To measure \( \kappa_{{xy}} \) accurately, any contamination from thermal conductivity \( \kappa_{{xx}} \) due to a slight misalignment of the two opposite transverse contacts is eliminated by field anti-symmetrization. This involves calculating \( \Delta T_{{y}}^{\mathrm{as}}(T,B) = \left[\Delta T_{{y}}(T,B) - \Delta T_{{y}}(T,-B) \right] / 2 \), where \( \Delta T_{{y}}^{\mathrm{as}}(T,B) \) represents the antisymmetrized \( \Delta T_{{y}}(T,B) \). The heat current along the \( {x}\) axis is generated by a strain gauge heater attached to one end of the sample, and the other end connected to a copper block using silver paint, serving as a heat sink.

For YBCO, which is orthorhombic, \( \kappa_{{yy}} \) was measured in a separate sample with the heat current applied along the \( y \) direction (corresponding to the crystallographic \( b \) axis) \cite{grissonnanche2016}. In this case, the thermal Hall conductivity \( \kappa_{{xy}} \) was calculated using the measured \( \kappa_{{yy}} \) in combination with the method described above.

For a more comprehensive discussion of the thermal transport measurement technique, the reader is referred to \cite{grissonnanche2016, grissonnanche2020,ashvini2024,chen2024,etienne2022, boulanger2020,chen2024}, where the measurements were carried out using the same experimental methodology.

\begin{figure*}
\centering
\hspace{-9.3mm} 
\includegraphics[width=0.7\textwidth]{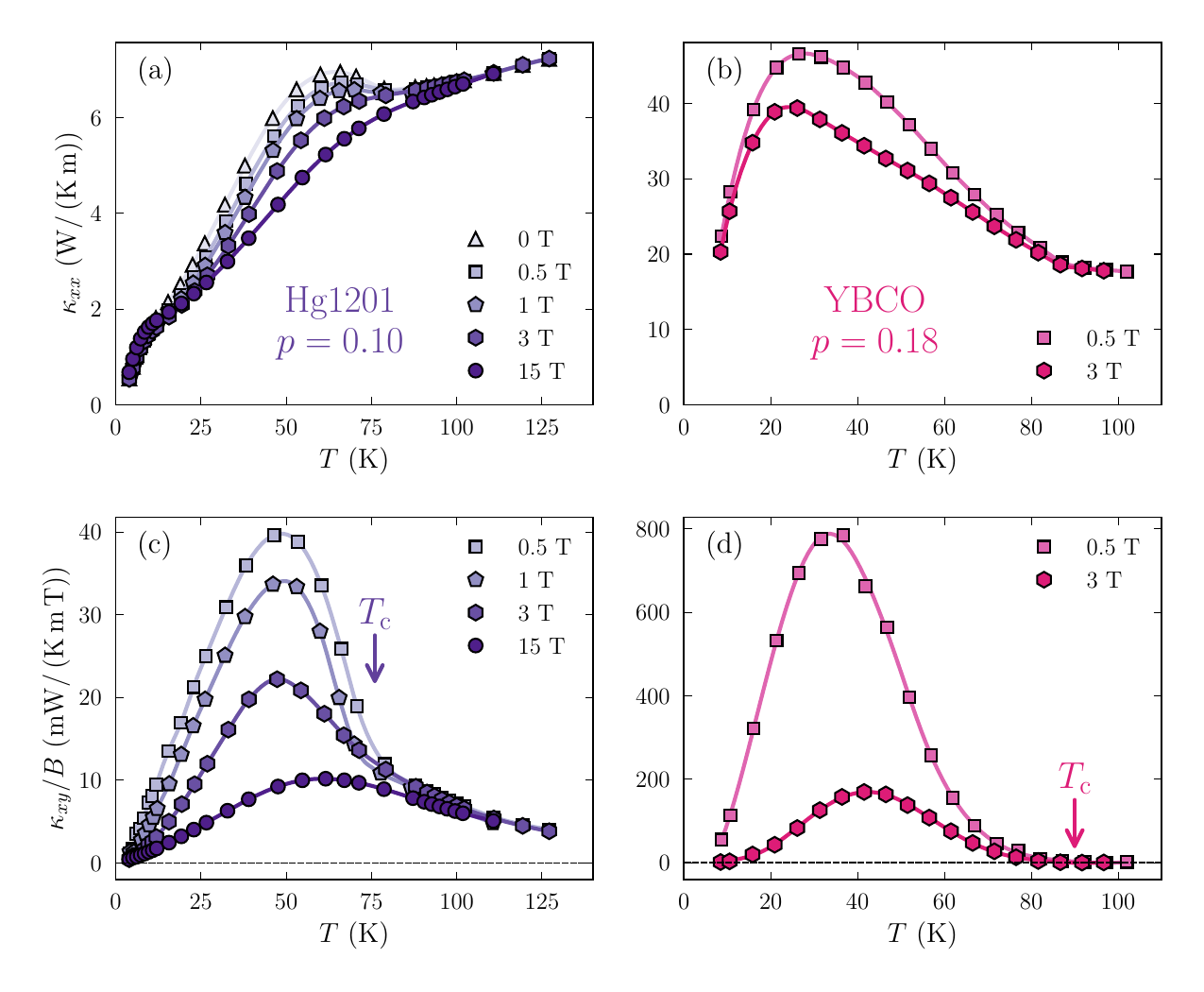}
\caption{\textbf{Thermal conductivities in Hg1201 and YBCO.} Longitudinal thermal conductivity \(\kappa_{{xx}}\) and transverse thermal conductivity normalized by field \(\kappa_{{xy}}/B\)  for Hg1201 (\(p = 0.10\)) and YBCO (\(p = 0.18\)) as a function of temperature \(T\). For Hg1201, data are shown for magnetic fields of 0 T (triangles), 0.5 T (squares), 1 T (pentagons), 3 T (hexagons), and 15 T (circles), with darker shades indicating higher fields. For YBCO, data are presented for 0.5 T (squares) and 3 T (hexagons). Lines serve as guides to the eye, while markers represent the data points. Panels (a)–(b) show \(\kappa_{{xx}}\): (a) Hg1201 (\(p = 0.10\), purple); (b) YBCO (\(p = 0.18\), pink). Panels (c)–(d) show \(\kappa_{{xy}}/B\). The arrows indicate the location of the superconducting transition.}
\label{Fig_Hg1201_YBCO_kxx_kxy}
\end{figure*}
\section{\label{sec:level3}RESULTS}

In Fig.~\ref{Fig_Hg1223_kxx_kxy}, the thermal conductivity $\kappa_{{xx}}$ and the thermal Hall conductivity divided by the magnetic field $\kappa_{{xy}} / B$ are presented for three Hg1223 samples at dopings $p = 0.09$, $0.10$, and $0.11$ for various magnetic fields ($B = 0$, $0.5$, $1$, $3$, and $15$ T) from $3$ K to $140$ K. 

The thermal conductivity $\kappa_{xx}$ exhibits a prominent peak upon entering the superconducting state at low fields in the \( p = 0.10 \) and \( p = 0.11 \) samples (top panels). This peak results from a sharp reduction in inelastic electron-electron scattering as electrons condense into Cooper pairs, leading to a substantial increase in the electronic mean free path \cite{Hirschfeld1996}. In contrast, the \( p = 0.09 \) sample shows a significantly reduced peak. 

When elastic scattering dominates over inelastic scattering, a peak in thermal conductivity is not expected in the superconducting state of a \( d \)-wave superconductor. The diminished peak in the $p=0.09$ sample [Fig.~1a] points to a larger ratio of elastic to inelastic scattering. This is in part due to the lower $T_{\mathrm{c}}$, making inelastic scattering at the transition weaker than in the samples with higher $T_{\mathrm{c}}$.
But it  may also reflect a higher level of disorder.

In the lower panels of Fig.~\ref{Fig_Hg1223_kxx_kxy}, we see that the magnitude of the peak in $\kappa_{{xy}} / B$ differs between the samples despite similar doping levels, where comparable electron densities are expected. This difference likely arises from variations in disorder, which we will quantify in the discussion section. The $p = 0.11$ sample has the highest peak value, namely $\kappa_{{xy}} / B \approx 800$ mW/(K·m·T) at $B = 0.5$ T, while the $p = 0.09$ sample shows a peak value of $\kappa_{{xy}} / B \approx 100$ mW/(K·m·T).

Fig.~\ref{Fig_Hg1201_YBCO_kxx_kxy} presents the thermal conductivity $\kappa_{{xx}}$ and the thermal Hall conductivity divided by the magnetic field $\kappa_{{xy}} / B$ for single-layer Hg1201 and bilayer YBCO at magnetic fields $B = 0.5$ T and $3$ T. For Hg1201, the same fields as used for Hg1223 are included.

The peak value of $\kappa_{{xy}} / B$ in Hg1201 is notably low compared to Hg1223, around $40$ mW/(K·m·T). This reduced conductivity is attributed to the higher disorder in Hg1201, as it lacks the multilayer structure that provides protection of the inner layer from impurities. In contrast, YBCO--known to be the cleanest among cuprates--exhibits a $\kappa_{{xy}} / B$ signal comparable to Hg1223 at $p = 0.11$, approximately $800$ mW/(K·m·T). This similar magnitude strongly suggests that Hg1223 is also exceptionally clean, with its inner plane protected by the outer planes, minimizing disorder.

\begin{figure}
\centering
\includegraphics[width=0.45\textwidth]{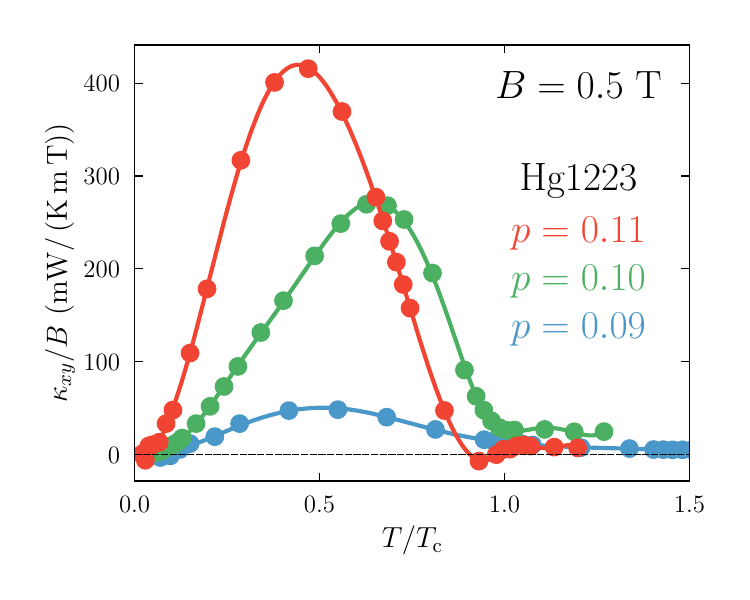}
\caption{Thermal Hall conductivity \(\kappa_{{xy}}\) as a function of reduced temperature \(T/T_{\mathrm{c}}\) for Hg1223 at \(B = 0.5\) T: \(p = 0.09\) (blue), \(p = 0.10\) (green), and \(p = 0.11\) (red). Lines serve as guides to the eye, while markers represent the data points. These data are used to extract the mean free path $\ell_{\mathrm{s}}$ in Hg1223.}
\label{Fig_Hg1223_kxy_0p5T}
\end{figure}

Fig.~\ref{Fig_Hg1223_kxy_0p5T} shows \(\kappa_{xy}/B\) measured in the lowest field for our three Hg1223 samples. 
These data are used to extract the mean free path of nodal quasiparticles, 
in the next section.

\section{\label{sec:level4}DISCUSSION}

The advantage of using thermal transport to estimate the mean free path of electrons is that it can be measured in the superconducting state, at low fields. However, to use the thermal conductivity $\kappa_{{xx}}$ is challenging because phonons also make a large contribution to $\kappa_{{xx}}$. This is why we turn to the thermal Hall conductivity $\kappa_{{xy}}$, which is typically dominated by electrons at low fields.

Before turning to the interpretation of the mean free path data, we first address the possible contribution of phonons to the thermal Hall signal in our samples. Since 2020, it has been established that phonons can generate a sizable thermal Hall response in both hole- and electron-doped cuprates \cite{grissonnanche2020,boulanger2020,lizaire2021,boulanger2022,boulanger2023}. In all such materials studied to date—including LSCO, Eu-LSCO, Nd-LSCO, Bi2201, NCCO, and PCCO—the phonon thermal Hall conductivity is consistently negative in sign.

Assuming the same sign applies to Hg1223, Hg1201, and YBCO, the fact that we observe a positive $\kappa_{{xy}}$ in all three compounds strongly suggests that the signal is electronic in origin. Furthermore, the magnitude of the thermal Hall conductivity in our measurements far exceeds typical phonon values. For instance, in LSCO ($x = 0.06$) at 50 K and 1 T, the phonon Hall conductivity is about $\kappa_{xy}/B \sim-2$ mW/(K·m·T) \cite{grissonnanche2019}, whereas the signal in Hg1223 (Fig.~\ref{Fig_Hg1223_kxx_kxy}) and YBCO (Fig.~\ref{Fig_Hg1201_YBCO_kxx_kxy}d) is roughly two orders of magnitude larger, and about one order of magnitude larger in Hg1201 (Fig.~\ref{Fig_Hg1201_YBCO_kxx_kxy}c). This shows that the phonon contribution to \(\kappa_{xy}\) is negligible in all samples studied here, at least at low fields.

\subsection{Model}

The model from \cite{Zhang2001} is first presented to derive the thermal mean free path from \(\kappa_{{xy}}\), and then extended for a broader application to any \textit{d}-wave superconductor using the theoretical framework from \cite{vekhter201}. In the weak magnetic field regime (\(\omega_{\mathrm{c}} \tau \ll 1\)), where \(\omega_{\mathrm{c}}\) is the cyclotron frequency, the Boltzmann transport formalism is valid. For this analysis, the thermal mean free path is extracted at \(B = 0.5~\text{T} \). This field value represents a compromise: it is high enough to yield a good signal-to-noise ratio while keeping \(\omega_{\mathrm{c}} \tau \ll 1\), ensuring the applicability of the theoretical framework. Moreover, the assumption \(\omega_{\mathrm{c}} \tau \ll 1\) will be verified later to ensure the self-consistency of this approach. Measuring \(\kappa_{{xy}}\) at low fields minimizes the phonon contribution, making the electronic response the dominant factor (see above).

The essence of this model is the assumption that, in the weak-field regime, the thermal Hall conductivity \(\kappa_{{xy}}\) is related to the longitudinal thermal conductivity of the quasiparticles \(\kappa_{\mathrm{qp}}\) through the following relation:

\begin{equation}\label{eq : Hall angle}
\kappa_{{xy}}/\kappa_{\mathrm{qp}} = \eta \omega_{\mathrm{c}} \tau = \eta \frac{\ell}{k_{\mathrm{F}} \ell_{B}^2},
\end{equation}

where \(\ell_{B} = \sqrt{\hbar / e B}\) is the magnetic length. The parameter \(\eta\) accounts for the anisotropy of the scattering path length across the Fermi surface \cite{Zhang2001}. However, due to limited knowledge of \(\eta\) in the studied samples, we assume \(\eta = 1\) for all three cuprates under investigation: Hg1223, YBCO and Hg1201.

Following Zhang et \textit{al}. \cite{Zhang2001} we define the thermal conductivity of nodal quasiparticles in the superconducting state :

\begin{equation}\label{eq : k_e}
\kappa_{\mathrm{qp}} = c_\text{qp} v_{\mathrm{F}} \ell / 4,
\end{equation}

where \(v_{\mathrm{F}}\) is the normal-state Fermi velocity in the nodal direction and \(c_\text{qp}\) is the specific heat of the nodal quasiparticles, which varies quadratically with temperature:

\begin{equation}\label{eq : specific heat}  
c_\text{qp} = \alpha_{\mathrm{c}} T^2.  
\end{equation}  

The prefactor \(\alpha_{\mathrm{c}}\) was measured experimentally in YBCO 
($p \simeq 0.16$)
as \(\alpha_{\mathrm{c}} = 0.064\) mJ/K\(^{3}\)·mol \cite{wright1999}. Combining the expressions for \(\kappa_{{xy}}/\kappa_{\mathrm{qp}}\) and \(\kappa_{\mathrm{qp}}\) leads to the following equation for the mean free path $\ell_{\mathrm{s}}$, derived in \cite{Zhang2001}:  

\begin{equation}\label{eq : mfp Ong}  
\ell_{\mathrm{s}} = 2 \ell_{B} \sqrt{\frac{\kappa_{{xy}} k_{\mathrm{F}}}{c_\text{qp} v_{\mathrm{F}} \eta}}.  
\end{equation}  

In addition to $k_{\mathrm{F}}$ and $v_{\mathrm{F}}$, this formula requires knowledge of the specific heat coefficient $\alpha_{\mathrm{c}}$, which is not often accessible. To address the practical limitations in accessing this parameter, we introduce an alternative formulation for $\ell_{\mathrm{s}}$ that depends instead on the gap velocity $v_{\Delta}$. This is achieved using the theoretical expression for $c_\text{qp}$ of \textit{d}-wave superconductors \cite{vekhter201}:

\begin{equation}\label{eq : specific heat new}  
c_\text{qp} = \frac{18 k_{\mathrm{B}}^3 \zeta(3)}{\hbar^2 d \pi v_{\mathrm{F}} v_\Delta}  T^2,  
\end{equation}

\begin{figure}
\centering
\includegraphics[width=0.45\textwidth]{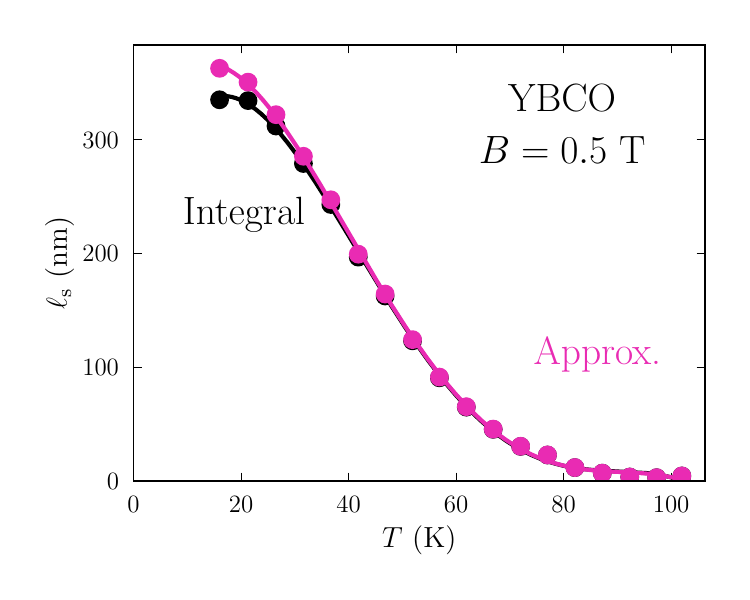}
\caption{Mean free path $\ell_{\mathrm{s}}$ as a function of temperature \(T\) in YBCO (\(p = 0.18\)), calculated using two different forms of the quasiparticle specific heat \(c_\mathrm{qp}\). The black markers represent $\ell_{\mathrm{s}}$ extracted using the integral form of \(c_\mathrm{qp}\), given in 
Eq.~(63) of \cite{vekhter201}, while the pink markers correspond to $\ell_{\mathrm{s}}$ obtained with the approximate expression  Eq.~(\ref{eq : specific heat new}). Lines are a guide to the eye. The two methods show excellent agreement for \(T > 30\) K, with deviations remaining below 10\% at lower temperatures. The uncertainty associated with this deviation is included in the error bar on the mean free path values we quote in the text and in Table~\ref{tab:sample_properties}.}
\label{Fig_mfp_approx_testing}
\end{figure}
where \(\zeta\) is the Riemann zeta function. A comparison between theoretical and experimental values of $\alpha_{\mathrm{c}}$ in YBCO shows excellent agreement—the theoretical estimate is only 1.2 times the measured value (using \(v_\mathrm{F}=2.5\times10^{5}\) m/s and \(v_\mathrm{\Delta}=1.5\times10^{4}\) m/s \cite{Vishik2014}). This supports the use of the theoretical form for analyzing Hg1223 and Hg1201, where experimental values of $\alpha_{\mathrm{c}}$ are not available.

Eq.~(\ref{eq : specific heat new}) is valid under the condition $E_\text{H} \ll T$, where $E_\text{H} \sim 30 \sqrt{B}$ K/T$^{1/2}$ represents the Doppler shift energy scale in \textit{d}-wave superconductors \cite{vekhter201}. For $B = 0.5$ T, $E_\text{H} \simeq 21$ K, making Eq.~(\ref{eq : specific heat new}) applicable above this temperature.

To validate the $T^2$ approximation for the quasiparticle specific heat [Eq.~(\ref{eq : specific heat new})], we compare in Fig.~\ref{Fig_mfp_approx_testing} the mean free path $\ell_{\mathrm{s}}$ obtained using two forms of $\alpha_{\mathrm{c}}$: the full integral expression from Eq.~(63) of Ref.~\cite{vekhter201}, valid across all temperatures and fields, and the simplified form from Eq.~(\ref{eq : specific heat new}) used in this work. As expected, both approaches yield nearly identical results at temperatures well above 21 K. At lower temperatures, where the approximation breaks down, the deviation in $\ell_{\mathrm{s}}$ is less than 10 \%; the associated uncertainty is included in the error bars of the quoted values (see Table~\ref{tab:sample_properties}).

Substituting Eq.~(\ref{eq : specific heat new}) into Eq.~(\ref{eq : mfp Ong}) yields:  

\begin{equation}\label{eq : mfp intermediate}  
\ell_{\mathrm{s}} = \ell_{B} \sqrt{\frac{2 \pi \hbar^2} {9 \zeta(3) k_{\mathrm{B}}^3 }}\sqrt{\frac{ v_{\Delta} k_{\mathrm{F}} d\kappa_{{xy}}} {\eta T^{2} }}.  
\end{equation}  

Note that this expression is broadly applicable to superconductors for which a Boltzmann transport approach is deemed valid, including multiband and three-dimensional cases, provided the Fermi surface and gap velocity are known.

\begin{figure}
    \centering
    \includegraphics[width=0.45\textwidth]{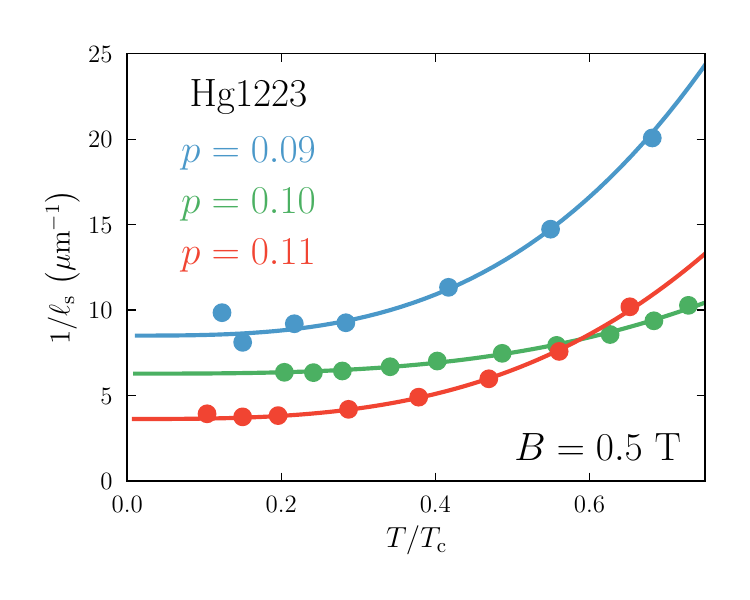}
    \caption{Inverse mean free path \(1/\ell_{\mathrm{s}}\) as a function of temperature \(T\) for Hg1223 at 0.5 T: \(p = 0.09\) (blue), \(p = 0.10\) (green), and \(p = 0.11\) (red). Markers represent the data points, while the lines correspond to \(T^3\) fits (\(a + b T^3\)). The residual value at $T = 0$ is given by \(\ell_{\mathrm{s0}} = 1 / a\), whose values are listed in Table~\ref{tab:sample_properties}.}
    \label{Fig_Hg1223_mfp.pdf}
\end{figure}
\subsection{Analysis}

From Eq.~(\ref{eq : mfp intermediate}), what we need is the value of \(k_\mathrm{F}\) and \( v_\Delta \), in the nodal direction (we assume $\eta = 1$.). This knowledge does not exist for all materials and all dopings. Detailed ARPES studies of the bilayer cuprate Bi2212 reveal that \( v_\Delta \) is constant and equal to \( 1.5 \times 10^4 \, \text{m/s} \) between \( p = 0.08 \) and \( p = 0.20 \) \cite{Vishik2012}. It is well known that the nodal wavevector varies only slightly between \( p = 0.09 \) and \( p = 0.20 \), with values close to \( k_\mathrm{F} = 0.74 \, \text{Å}^{-1} \) \cite{fujita2014}. ARPES data on Hg1201 yield similar values of \(k_{\mathrm{F}}\) and \(v_{\mathrm{F}}\) \cite{Vishik2014}.

A measurement of the zero-field thermal conductivity in the \( T \to 0 \) limit yields a residual linear term \( \kappa_{0} / T \) which is independent of disorder (i.e., universal) \cite{Taillefer1997} and given only by the ratio \( v_\mathrm{F} / v_\Delta \) \cite{Durst2000}. A prior measurement on YBCO \( p = 0.18 \) gave \( \kappa_{0} / T = 0.16 \, \text{mW/K}^2\text{cm} \) \cite{hill2000}, in excellent agreement with the ARPES values for \( v_\mathrm{F} \) and \( v_\Delta \) in Bi2212, namely \( v_\mathrm{F} = 2.5 \times 10^5 \, \text{m/s} \) and \( v_\Delta = v_\mathrm{F} / 16 \) \cite{Vishik2010, Vishik2012}, as discussed in \cite{Reid_2012}. We are therefore fully justified to use \( k_\mathrm{F} = 0.74 \, \text{Å}^{-1} \) and \( v_\Delta = 1.5 \times 10^4 \, \text{m/s} \) in Eq.~(\ref{eq : mfp intermediate}) for YBCO \( p = 0.18 \). Given the weak doping dependence of these two parameters (even in the presence of the pseudogap and charge order), we will also use the same values of \(k_{\mathrm{F}}\) and \(v_{\mathrm{F}}\) for Hg1223 (\( p = 0.09, 0.10, 0.11 \)).

Eq.~(\ref{eq : mfp intermediate}) was used to extract \(1 / \ell_{\mathrm{s}}\) from \(\kappa_{{xy}}\) for the three Hg1223 samples at \(B = 0.5\) T (see Fig.~\ref{Fig_Hg1223_kxy_0p5T}), and the results are plotted in Fig.~\ref{Fig_Hg1223_mfp.pdf}. The inverse thermal mean free path follows a cubic temperature dependence, consistent with the expected behavior of the scattering rate \((1/\tau_{\mathrm{s}} = v_{\mathrm{F}} /\ell_{\mathrm{s}} )\). A fit to the function \(1/\ell_{\mathrm{s}} = a + bT^3\) was performed for each sample, where the parameter \(a\) represents the elastic scattering of nodal quasiparticles on impurities, allowing the extraction of the residual mean free path \(\ell_{\mathrm{s0}}\) later in this work. The coefficient \(b\) characterizes the strength of inelastic scattering processes.  

An intriguing feature in Fig.~\ref{Fig_Hg1223_mfp.pdf} is the significant difference in slopes between the three dopings, corresponding to variations in \(b\). The extracted values of \(b\) are 79, 11, and 16 (in units of 1/m·K³) for Hg1223 samples with \(p = 0.09\), \(p = 0.10\), and \(p = 0.11\), respectively. 
Theoretical studies have predicted a cubic temperature dependence of the nodal quasiparticle scattering rate in clean \(d\)-wave superconductors \cite{walker2000, dahm2005, duffy2001,Quinlan1994}, with quasiparticle-quasiparticle interactions naturally giving rise to this behavior via Fermi’s Golden Rule \cite{walker2000}. 
It has also been shown that scattering of nodal quasiparticles 
by spin fluctuations can lead to a \(T^3\) dependence \cite{duffy2001, dahm2005,Quinlan1994}.

The extensive theoretical literature supporting spin-fluctuation scattering as the origin of the \(T^3\) dependence, combined with the fact that the sample closest to the antiferromagnetic phase (\(p = 0.09\)) exhibits the highest \(b\) value, suggests that spin fluctuations could be the primary mechanism.

Rather than drawing a definitive conclusion, we present these results as a foundation for future investigations into the precise origin of the observed \(T^3\) dependence.

\begin{figure}
\centering
\includegraphics[width=0.45\textwidth]{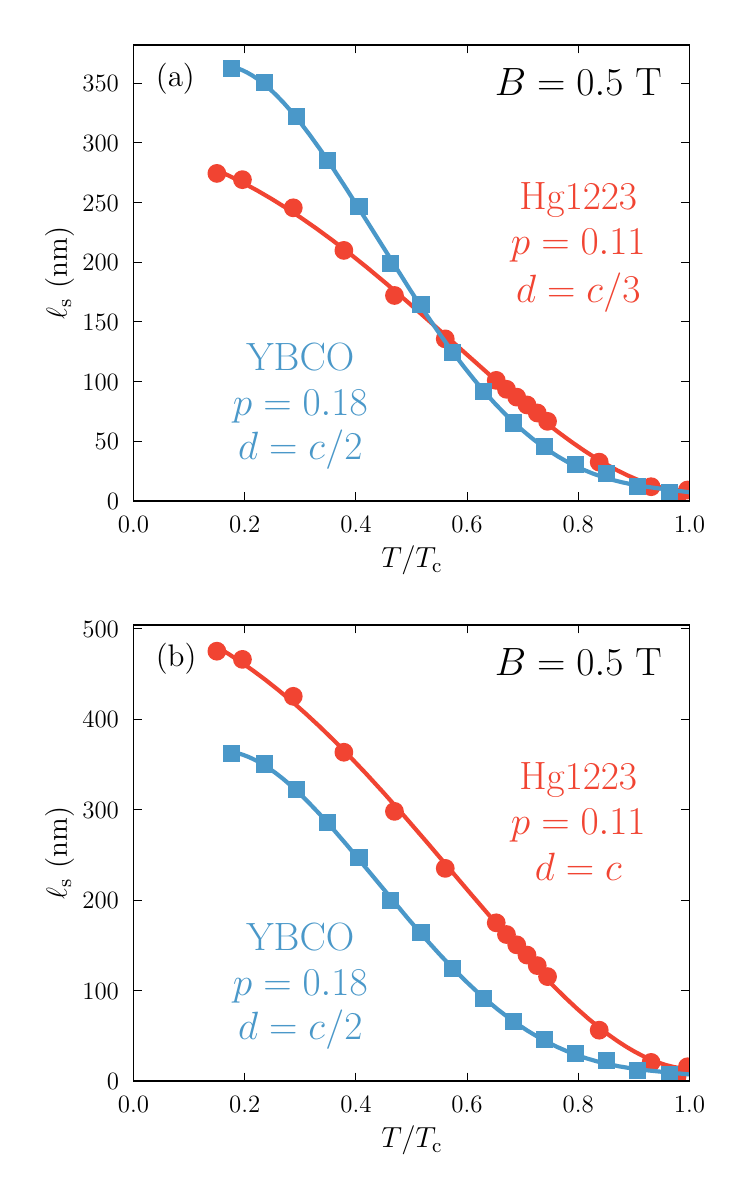}
\caption{Mean free path $\ell_{\mathrm{s}}$ as a function of reduced temperature \(T/T_{\mathrm{c}}\) for Hg1223 (\(p = 0.11\), red, circle) and YBCO (\(p = 0.18\), blue, square) at a magnetic field of 0.5 T. Markers represent the data points, and lines serve as guides to the eye.
a) Assuming that the three planes in Hg1223 contribute equally ($d=c/3$);
b) assuming that only the inner plane contributes ($d=c$).}
\label{Fig_mfp_Hg1223_vs_YBCO}
\end{figure}

For Hg1223, the residual mean free paths in the superconducting state are \(\ell_{\mathrm{s0}} = 1180 \, \pm \, 170 \, \text{\AA}\), \(1590 \, \pm \, 230 \, \text{\AA}\), and \(2760 \, \pm \, 390 \, \text{\AA}\) for \(p = 0.09, 0.10\) and 0.11, respectively. 
These values of \(\ell_{\mathrm{s0}}\) correspond to \(1 / a\), where \(a\) is the parameter extracted from the cubic fit \(1 / \ell = a + bT^3\) applied to the data in Fig.~\ref{Fig_Hg1223_mfp.pdf} described earlier. 
These values are significantly higher than those typically observed in the normal state of clean cuprates such as Tl2201, where \(\ell_{\mathrm{n0}}\) is roughly \(500 \, \text{\AA}\) \cite{Mackenzie1996}.

In Fig.~\ref{Fig_mfp_Hg1223_vs_YBCO}, we compare the mean free path derived from Eq.~(\ref{eq : mfp intermediate}) for data on Hg1223 $p=0.11$ and data on YBCO $p=0.18$, using the same \(k_{\mathrm{F}}\) (\(0.74 \, \text{Å}^{-1})\)  and \(v_{\mathrm{\Delta}}\) (\(1.5\times10^{4} \, \text{m/s})\). The comparable values of \(\ell_{\mathrm{s0}} = 2760 \, \pm \, 390 \, \text{\AA}\) for Hg1223 and \(\ell_{\mathrm{s0}} = 3530 \, \pm \, 500 \,\text{\AA}\) for YBCO highlight the exceptional cleanliness of Hg1223, as YBCO is the cleanest amongst cuprates. Furthermore, the value of \(\ell_{\mathrm{s0}} = 2760 \, \pm \, 390 \, \text{\AA}\) for Hg1223 at \(p = 0.11\) is significantly higher than that of \(1040 \, \pm \, 150 \, \text{\AA}\) for Hg1201. This suggests that the mean free path in the normal state of Hg1223 is likely much larger than the \(200 \, \text{\AA}\) reported for Hg1201 from the Dingle temperature of quantum oscillations \cite{chan2016b}. This observation is consistent with reports of quantum oscillations in all three compounds (YBCO, Hg1201 and Hg1223) \cite{oliviero2022,oliviero2024,barisic2013a,doiron-leyraud2007,chan2016b}.

When comparing the mean free path of Hg1223 with that of the cleanest cuprate, fully oxygenated YBCO, the values are similar (see Fig.~\ref{Fig_mfp_Hg1223_vs_YBCO}(a)). 
However, the residual mean free path for Hg1223 represents a lower bound, as we assumed an equal contribution from all three layers. 
This assumption directly affects the value of the interplane separation \(d\), which is taken as \(d = 5.3 \, \text{\AA}\) when all three planes contribute. 
If only one plane contributed, the interplane separation increases to \(d = 16 \, \text{\AA}\), leading to a higher calculated mean free path of \(\ell_{\mathrm{s0}} \approx 4600 \, \text{\AA}\) compared to \(\ell_{\mathrm{s0}} = 2760 \, \text{\AA}\) for \(d = 5.3 \, \text{\AA}\)
(see Fig.~\ref{Fig_mfp_Hg1223_vs_YBCO}(b)) 
This suggests that the true mean free path of Hg1223 likely falls between these two extremes, indicating that its sample quality could exceed that of the cleanest YBCO. Hg1223, being the strongest cuprate superconductor, is also likely the cleanest.

\begin{figure}
\centering
\includegraphics[width=0.45\textwidth]{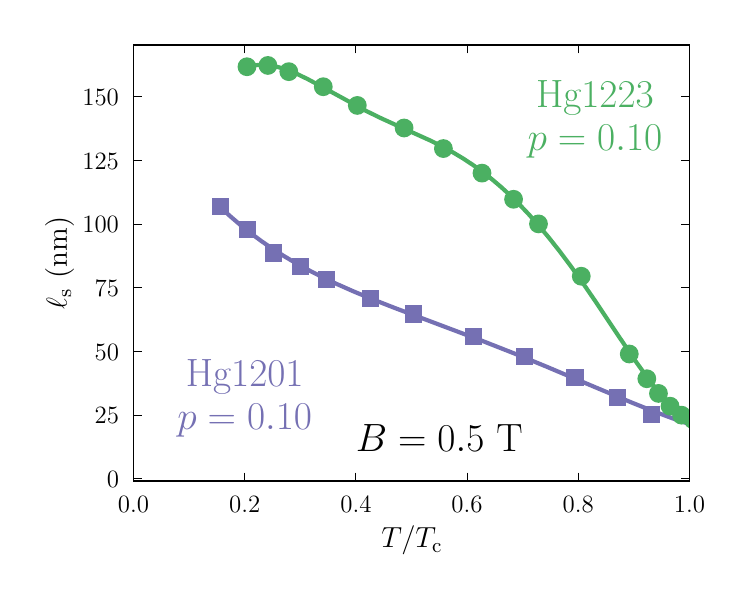}
\caption{Mean free path $\ell_{\mathrm{s}}$ as a function of reduced temperature \(T/T_{\mathrm{c}}\) for Hg1223 (\(p = 0.10\), green, circle) and Hg1201 (\(p = 0.10\), purple, square). Markers represent the data points, and lines serve as guides to the eye.
From Eq.~(\ref{eq : mfp intermediate}) using the same values of \(k_{\mathrm{F}}\) and \(v_{\mathrm{\Delta}}\), and assuming that the three planes in Hg1223 contribute equally ($d=c/3$).}

\label{Fig_mfp_Hg1223_vs_Hg1201}
\end{figure}

To better understand the role of the outer layers in the sample quality of Hg1223, we compare its mean free path with that of its single-layered counterpart, Hg1201, both doped with \(p = 0.10\). The comparison shows that trilayered Hg1223 exhibits significantly higher sample quality than single-layer Hg1201 (see Fig.~\ref{Fig_mfp_Hg1223_vs_Hg1201}). Once again, the estimated mean free path in Hg1223 represents a lower bound, highlighting the protective effect of the outer layers on the inner plane.

To validate the self-consistency of our framework, we verify the hypothesis \(\omega_{\mathrm{c}} \tau \ll 1\) for the YBCO sample, which has the highest residual mean free path, \(\ell_{\mathrm{s0}} = 3530 \, \text{\AA}\). Using the expression \(\omega_{\mathrm{c}} \tau = e B \ell_{\mathrm{s0}} / \hbar k_{\mathrm{F}}\), we estimate \(\omega_{\mathrm{c}} \tau \sim 0.04\) at \(B = 0.5 \, \text{T}\), confirming the self-consistency of our analysis.

\subsection{Field dependence}

\begin{figure}
\centering
\includegraphics[width=0.45\textwidth]{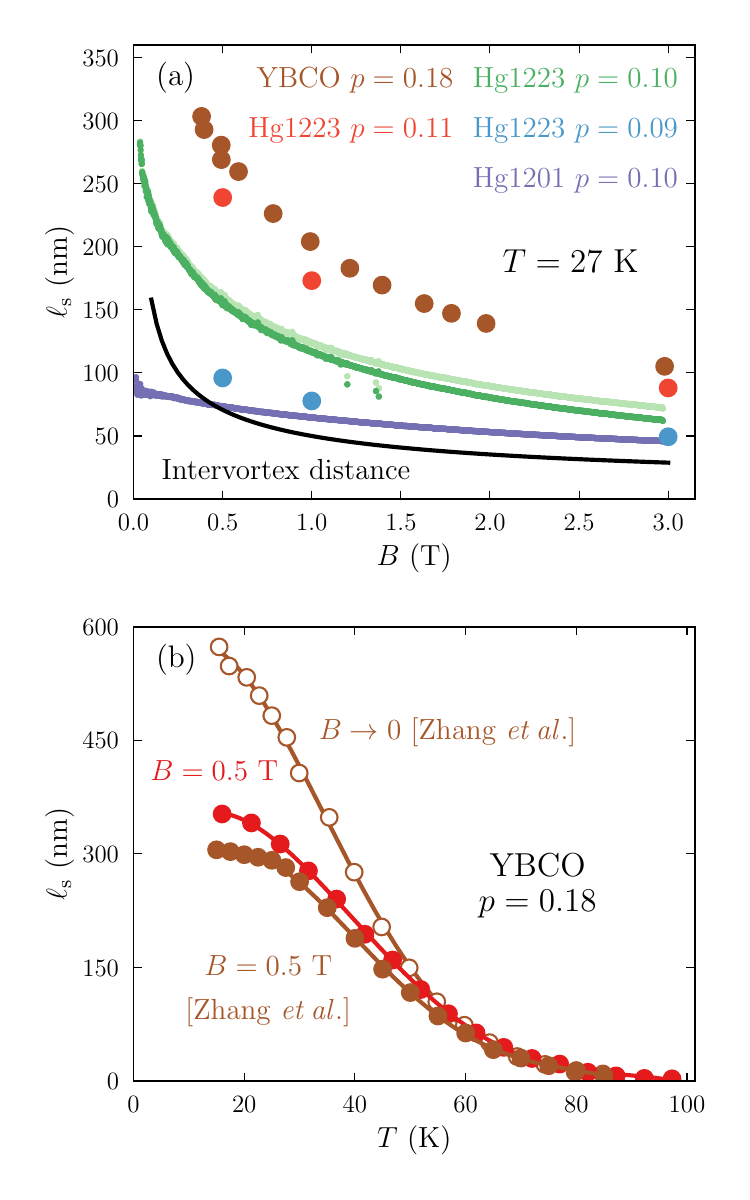}
\caption{(a) Mean free path $\ell_{\mathrm{s}}$ as a function of magnetic field \(B\) at \(T = 27\) K for YBCO (\(p = 0.18\), brown), Hg1223 (\(p = 0.09\), blue; \(p = 0.10\), green; \(p = 0.11\), red), and Hg1201 (\(p = 0.10\), purple). The YBCO data at \(B = 0.5\) T are taken from Fig.~2 of \cite{Zhang2001}. For Hg1201 (\(p = 0.10\)) and Hg1223 (\(p = 0.10\)), field sweeps of \(\kappa_{{xy}}\) were performed. For Hg1223 (\(p = 0.09\)) and (\(p = 0.11\)), \(\ell_{\mathrm{s}}\) was extracted from temperature step measurements of \(\kappa_{{xy}}\) at \(B = 0.5\), 1, and 3 T (Fig.~\ref{Fig_Hg1223_kxx_kxy}). The black line represents the intervortex distance, given by \(50 / \sqrt{B}\) nm \cite{vekhter201}. The mean free path shown here is calculated using the general, $H$- and $T$-dependent expression for the specific heat from Ref.~\cite{vekhter201}, except for the light green curve (\(p = 0.10\)) where Eq.~(\ref{eq : mfp intermediate}) was used to illustrate the difference between the two approaches. (b) Mean free path $\ell_{\mathrm{s}}$ as a function of temperature \(T\) for YBCO (\(p = 0.18\)). The red curve corresponds to \(\ell_{\mathrm{s}}\) extracted from our data at \(B = 0.5\) T, while the brown curves represent the data from \cite{Zhang2001} at \(B = 0.5\) T (full circles) and in the \(B \to 0\) limit open circles).}

\label{Fig_mfp_field_dependence}
\end{figure}

The mean free path $\ell_{\mathrm{s}}$ of nodal quasiparticles in the superconducting state depends on magnetic field, for two reasons.
First, inelastic scattering grows with field, since the field excites quasiparticles and hence electron-electron scattering.
Secondly, vortices can scatter heat carriers, including quasiparticles. 
This is why in clean samples the upper critical field $H_{c2}$ -- the field below which vortices form -- can be detected by measuring the thermal conductivity:
$\kappa_{xx}$ drops sharply as $H$ is reduced below $H_{c2}$, at $T \simeq 0$ \cite{Grissonnanche2014}.
In Fig.~\ref{Fig_mfp_field_dependence}(a), we show the field dependence of $\ell_{\mathrm{s}}$ in the 5 samples investigated here.
The calculation uses the general, $H$-and $T$-dependent expression for the specific heat from Ref.~\cite{vekhter201}. (Note that using Eq.~(\ref{eq : mfp intermediate}) instead makes little difference.)
In two cases, Hg1223 $p=0.10$ and Hg1201, a continuous field sweep was performed.
In Hg1223 $p=0.10$, $\ell_{\mathrm{s}}$ exhibits a strong field dependence, increasing from 160 nm at \(B = 0.5\) T to approximately 280 nm at \(B \to 0\).
A similarly strong $H$ dependence is observed in Hg1223 $p=0.11$ and YBCO. 
In contrast, Hg1201 shows a much weaker field dependence, with $\ell_{\mathrm{s}}$ increasing only slightly from 70 nm at \(B = 0.5\) T to 90 nm at \(B \to 0\), comparable to what is seen in Hg1223 $p=0.09$.

This field dependence implies that an estimate of $\ell_{\mathrm{s}}$ should be taken in the \(B \to 0\) limit. But comparing different samples or different cuprates, at a given finite field is nonetheless reasonable.

In Fig.~\ref{Fig_mfp_field_dependence}(b), we reproduce the data of Zhang {\it et al.} both at $B = 0.5$~T and at $B \to 0$ \cite{Zhang2001}.
We see that their data on YBCO $p=0.18$ are in excellent quantitative agreement with ours, when both are taken at $B = 0.5$~T. We also see that the $T \to 0$ value is roughly twice as large at $B \to 0$.

\section{\label{sec:Summary}CONCLUSION}

We have measured the thermal Hall conductivity of three high-$T_{\mathrm{c}}$ cuprate superconductors, with a focus on the trilayer material Hg1223. From our data, we extract the quasiparticle mean free path in the superconducting state via a simple model. By comparing the results with YBCO and Hg1201, we find that Hg1223 exhibits exceptional sample quality, potentially surpassing even the cleanest YBCO, the gold standard among cuprates. The trilayer structure of Hg1223 may play a crucial role, with the outer layers protecting the inner plane from disorder, resulting in a significantly higher mean free path compared to single-layer Hg1201. This suggests that the strongest cuprate superconductor, Hg1223, is also the cleanest.

We observe a cubic temperature dependence in the inverse mean free path, a behavior predicted for clean \textit{d}-wave superconductors. More work is needed to understand how the strength of the inelastic scattering responsible for this $T^3$ dependence varies with doping.

\begin{acknowledgments}

We thank Seyed Amirreza Ataei, Lu Chen, Robin Durand, Koko Mino and Juan Vieira Giestinhas for fruitful discussions. We also thank Simon Fortier for his assistance with the experiments, and the cryogenics team at the Institut Quantique for their support. We are grateful to the Centre Tara for hosting a writing retreat that contributed to the completion of this manuscript. L. T. acknowledges support from the Canadian Institute for Advanced Research (CIFAR) as a CIFAR Fellow and funding from the Institut Quantique, the Natural Sciences and Engineering Research Council of Canada (Grant No.~PIN:123817), the Canada Foundation for Innovation, and a Canada Research Chair. C.P. acknowledges  support from the EUR grant NanoX No.~ANR-17-EURE-0009 and from the ANR grant NEPTUN No.~ANR-19-CE30-0019-01. G.G. acknowledges support from STeP2 No.~ANR-22-EXES-0013, QuantExt No.~ANR-23-CE30-0001-01, Audace CEA No.~ANR-24-RRII-0004 and the École Polytechnique Foundation. M. Altangerel, Q. Barthélemy, É. Lefrançois, J. Baglo, M. Mezidi, A. Vallipuram, E. Campillo and L. Taillefer have benefitted from their affiliation to the RQMP \cite{funding}.  This research was undertaken thanks, in part, to funding from the Canada First Research Excellence Fund. This research project No.~324046 is made possible thanks to funding from the Fonds de recherche du Québec.

\end{acknowledgments}

\section*{DATA AVAILABILITY}

The data that support the findings of this article are not publicly available. The data are available from the authors upon reasonable request.

\nocite{*}

\bibliography{bibliography}

\end{document}